\documentclass[1, 12pt]{elsarticle}
\usepackage[english]{babel}
\usepackage{amsmath}
\usepackage[toc,page]{appendix}
\usepackage{amssymb}        
\usepackage{amsthm}        
\usepackage{natbib}         
\usepackage{mathtools}
\usepackage{graphicx}       
\usepackage{subfigure}      
\usepackage{color}
\usepackage{gensymb} 
\journal{Acta Materialia}

\title{An atomistic investigation of the interaction of dislocations with Guinier-Preston zones in Al-Cu alloys} 

\author{G. Esteban-Manzanares$^{1, 2, 3}$}
\author{E. Mart{\'\i}nez$^{3}$}
\author{J. Segurado$^{1, 2}$}
\author{L. Capolungo$^{3}$}
\author{J. LLorca$^{1, 2, }$\corref{cor1}}
\address{$^1$ IMDEA Materials Institute, C/ Eric Kandel 2, 28906, Getafe, Madrid, Spain. \\  $^2$ Department of Materials Science, Polytechnic University of Madrid/Universidad Polit\'ecnica de Madrid, E. T. S. de Ingenieros de Caminos. 28040 - Madrid, Spain. \\  $^3$ Material Science and Technology Division, MST-8, Los Alamos National Laboratory, Los Alamos 87545 NM, USA.}

\cortext[cor1]{Corresponding author}

\begin{document} 

\begin{abstract}

The interaction between edge dislocations and Guinier-Preston zones in an Al-Cu alloy was analyzed by means of atomistic simulations. The different thermodynamic functions that determine the features of these obstacles for the dislocation glide were computed using molecular statics, molecular dynamics and the nudged elastic band method. It was found that Guinier-Preston zones are sheared by dislocations and the rate at which dislocations overcome the precipitate is controlled by the  activation energy, $\Delta U$, in agreement with the postulates of the harmonic transition state theory. Moreover, the entropic contribution to the Helmholtz activation free energy was in the range 1.3-1.8$k_b$, which can be associated with the typical vibrational entropy of solids.   Finally, an estimation of the initial shear flow stress as a function of temperature was carried out from the thermodynamic data provided by the atomistic simulations. Comparison with experimental results showed that the effect of the random precipitate distribution and of the dislocation character and dislocation/precipitation orientation has to be taken into account in the simulations to better reproduce experiments.

\end{abstract}


\maketitle 

\newpage
\section{Introduction} \label{Intro}

Dislocations are the main carriers of plastic deformation in metallic materials. Since the critical shear stress to promote dislocation glide  in the most suitable crystallographic planes in most metals is very low, one of the main goals of physical metallurgy has been to increase the strength of metals. One strategy to achieve this objective is to insert different types of obstacles in the dislocation glide plane to hinder dislocation glide. Grain boundaries can act as such obstacles in polycrystalline samples while forest dislocations that develop during deformation as a result of dislocation multiplication also impede dislocation glide. The presence of alloying elements in a solid solution and of precipitates also lead to strengthening, termed as solution hardening and precipitation hardening, respectively. Precipitation hardening is more efficient than solute hardening to increase the yield strength of metallic alloys \citep{KN63} and it is always present in high strength metallic alloys. Moreover, since the energy barrier to dislocation glide induced by precipitates is much higher than that due to solutes, precipitate hardening is still effective at high temperatures while solute-hardening decreases rapidly with temperature \citep{KN63, VLG17}.

The dislocation-precipitate interaction mechanism depends on the morphological details of matrix and precipitate. Large precipitates, normally incoherent with respect to the matrix, are usually considered impenetrable such that dislocations overcome the obstacle by the formation of an Orowan loop \citep{Orowan1948}. This process is controlled by the line tension of the dislocation and can be analyzed with continuum models (e.g. discrete dislocation dynamics) that can take into account the influence of the elastic mismatch, precipitate shape and orientation, eigenstrains, etc. on the critical resolver shear stress to overcome the precipitate \citep{XSC04, QB09,  MND11, GFM15, ZSD17, SEP18}. On the contrary, small precipitates are usually coherent and tend to be sheared by dislocations. Quantifying their effects on strength is more complex as several factors have to be considered simultaneously including the energy associated with the new interface area created when the dislocation shears the precipitate, the stresses induced by the lattice mismatch and, in specific cases, the strengthening due to breakage of the long-range order in the matrix or the precipitate \citep{N97}. 

Continuum models can still be used to simulate precipitate strengthening in the case of shearable precipitates from a phenomenological perspective by increasing the lattice resistance or decreasing the dislocation mobility within the precipitate \citep{HZT12, GFM15, VDR09, HRU17}. However, both parameters have to be calibrated and it is not clear that the continuum hypotheses can always be applied in this context. Furthermore, classical atomistic simulations can account for the physical mechanisms of dislocation-precipitate interactions at this length scale and have been used for this purpose using the parallel array of dislocations strategy \citep{osetsky2003atomicI}. The critical resolved shear stress (CRSS) needed to overcome the precipitate can be computed in these simulations within the framework of Molecular Statics (MS) or Molecular Dynamics (MD) simulations, which have been widely used to analyze the interaction of dislocations with radiation defects as well as nm-sized voids and precipitates \citep{HSC00, SW10, proville2010dislocation, BTM11, LLH14, LGL16, bahramyan2016molecular, yanilkin2014dynamics}.

The limitation of these approaches is that they only provide the CRSS at 0K (MS) or at finite temperatures and very high strain rates (MD) but cannot  directly address the finite temperature/quasi-static strain rate regime. Moreover, the dislocation-precipitate interaction is a thermally activated process \citep{kocks1975thermodynamics}  and -- following transition state theory  (TST) \citep{GLE41} -- accurate predictions of rate and thermal effects can only be achieved if the free energy barrier (rather than the CRSS) is precisely determined. This strategy has been used to analyze dislocation-mediated phenomena in crystalline solids, such as dislocation cross-slip \citep{VRL00, WC09} and dislocation nucleation \citep{HL10, NBW11}. More recently, different variants of TST, including its harmonic approximation \citep{vineyard1957frequency} coupled to path sampling techniques to obtain energetic barriers \citep{saroukhani2016harnessing, SW17}  have been employed to determine the rate at which dislocations overcome either precipitates or solute atoms. 

The free energy barrier has to be evaluated to compute the activation rate using TST. This can be achieved by means of direct MD simulations or thermodynamic integration \citep{straatsma1988free} although the associated computational cost is very high. However, if harmonic transition state theory (HTST) holds, the rate can be obtained from the activation enthalpy barrier, which might be calculated using multireplica static procedures, such as the nudged elastic band (NEB) method \citep{jonsson1998nudged,henkelman2000climbing} or similar approaches \citep{henkelman1999dimer,barkema1996event,malek2000dynamics}, with much less computational cost. 

It is worth mentioning that the number of degrees of freedom in a system containing dislocations and precipitates is large, leading to very rough energy landscapes with many local minima. Hence, it is not guaranteed that the initial and final configurations used in multireplica simulations (which are usually obtained with MS) are representative of the actual physical mechanisms. For instance, MS simulations of the interaction of dislocations with Guinier-Preston (GP) zones in Al-Cu alloys predict the formation of Orowan loops around the precipitate \citep{SW10}, while precipitate shearing is expected for this coherent precipitate according to theory and experiments \cite{BFK61, BL18}. To overcome this limitation, an approach based on the combination of MS and MD simulations is presented in this paper to determine the initial and final configurations of the interaction of an edge dislocation with a GP zone in Al-Cu alloys. This strategy is used along with MD simulations to determine the main thermodynamic parameters within the framework of TST to determine the rate at which the dislocation overcomes the GP. This information is used to estimate the flow stress of an Al-Cu alloy reinforced with GP zones.

The paper is organized as follows. Transition state theory and its harmonic approximation is presented in section \ref{sec2:TST}. The  framework of the atomistic simulations (including MS, MD as well as the NEB method) is given in section \ref{sec3:Proc}, while the simulation results, the identification of the thermodynamic parameters and the discussion can be found in section \ref{sec4:Therm}. A prediction of the flow stress based on the atomistic results is presented in section \ref{sec5:FlowStress} and  the main conclusions of this investigation are summarized in section \ref{sec6:Conclusion}. 

\section{Transition state theory} \label{sec2:TST}

TST establishes that the rate $\Gamma$ at which an ergodic system crosses a dividing surface  is given by
\begin{equation}
\Gamma =\nu\: \exp{\left(\frac{-\Delta G(\tau)}{k_b T}\right)}
\label{E1}
\end{equation}

\noindent where $\nu$ is the fundamental attempt frequency,  $k_b$ the Boltzmann constant, $T$ the absolute temperature  and $\Delta G$ the activation Gibbs free energy that has to be supplied by thermal fluctuations to cross the dividing surface \citep{GLE41}. In the case of a dislocation that tries to overcome an obstacle under the action of an applied stress $\tau$, work is performed by applied stress to assist the transition by reducing the energy barrier. The activation Gibbs free energy necessary to overcome the obstacle can be expressed as 

\begin{equation}
\Delta G(\tau) =\Delta H (\tau) - T \Delta S_\tau (\tau, T)
\label{E2}
\end{equation}

\noindent where $\Delta H (\tau)$ is the activation enthalpy, which is a function of the applied stress, and  $\Delta S_\tau (\tau, T)$ the activation entropy \citep{kocks1975thermodynamics}. The subindex $\tau$ in the activation entropy denotes that it was obtained under constant stress. 

If the dislocation tries to overcome the obstacle at constant applied strain $\gamma$, the rate $\Gamma$ can be expressed as 

\begin{equation}
\Gamma =\nu\: \exp{\left(\frac{-\Delta F(\gamma)}{k_b T}\right)}
\label{E3}
\end{equation}

\noindent where $\Delta F(\gamma)$ is the Helmholtz activation free energy. $\Delta F(\gamma)$ stands for the energy barrier imposed by the obstacle in absence of external work and it is given by 

\begin{equation}
\Delta F(\gamma) =  \Delta U (\gamma) - T \Delta S_\gamma (\gamma, T)
\label{E4}
\end{equation}

\noindent where $\Delta U (\gamma)$ stands for the activation energy, which depends on the applied strain, and $\Delta S_\gamma(\gamma, T)$ is the activation entropy at constant strain, which may depend on the applied strain and temperature. 

It has been shown that $\Delta G(\tau) \approx \Delta F(\gamma)$  when the volume of the crystal is much larger than the activation volume of the thermally activated process  and $\tau$ and $\gamma$ are conjugate variables \cite{SR11}. Thus, both eqs. \eqref{E1} and \eqref{E3} can be used to determine the rate at which the obstacle is overcome, which should not be dependent on whether the crystal is subjected to a constant stress or to a constant strain that corresponds to the same stress.

A consequence of this result is that activation entropy at constant strain ($\Delta S_\gamma$) and at constant stress ($\Delta S_\tau$) are not the same, and they are related according to \cite{SR11}

\begin{equation}
\left. \Delta S_\tau- \Delta S_\gamma=- V_0 \frac{\partial \tau}{\partial T}\right|_\gamma
\label{E5}
\end{equation}

\noindent where $V_0$ is the activation volume.

Following these thermodynamic relations, the rate $\Gamma$ can be expressed as

\begin{equation}
\Gamma =  \nu\: \exp{\left(\frac{\Delta S_\gamma(\gamma, T)}{k_b}\right)} \exp{\left(-\frac{\Delta U(\gamma)}{k_b T}\right)}
\label{E6}
\end{equation}

\noindent This expression can be simplified if the harmonic transition state theory (HTST) approximation is applicable to the particular phenomenon under study \citep{V57, Hara2010}. HTST assumes that the entropy of the system does not depend on the temperature, and is equal to the ratio between the product of the normal frequencies in the initial stable configuration and in the saddle point configuration. Thus,  the rate can be written as 

\begin{equation}
\Gamma _{HTST} = \nu_{eff} \: \exp{\left(-\frac{\Delta U (\gamma)}{k_b T}\right)} 
\label{E7}
\end{equation}
 
\noindent where  $\nu_{eff}$ is the effective attempt frequency, which depends on the fundamental attempt frequency, $\nu$, and on the entropic factor $\exp{\left(\Delta S_\gamma/k_b\right)}$ \citep{granato1964entropy, sobie2017modal}. Thus, the HTST rate at which dislocations overcome the obstacle depends only on the activation energy and the effective attempt frequency according to eq. (\ref{E7}). All these parameters can be determined by means of atomistic simulations.

The activation Helmholtz free energy $\Delta F(\gamma)$ can be computed as a function of the applied strain,  as well as the fundamental attempt frequency ($\nu$), calculating the rate to overcome an obstacle using MD simulations. In addition, the activation energy $\Delta U(\gamma)$ can be evaluated using the NEB method. This information can be used to estimate the entropic contribution as a function of the applied strain and temperature and to establish the applicability of HTST to this particular case.

\section{Atomistic simulations} \label{sec3:Proc}

\begin{figure}[h]
\centering
\includegraphics[scale=0.55]{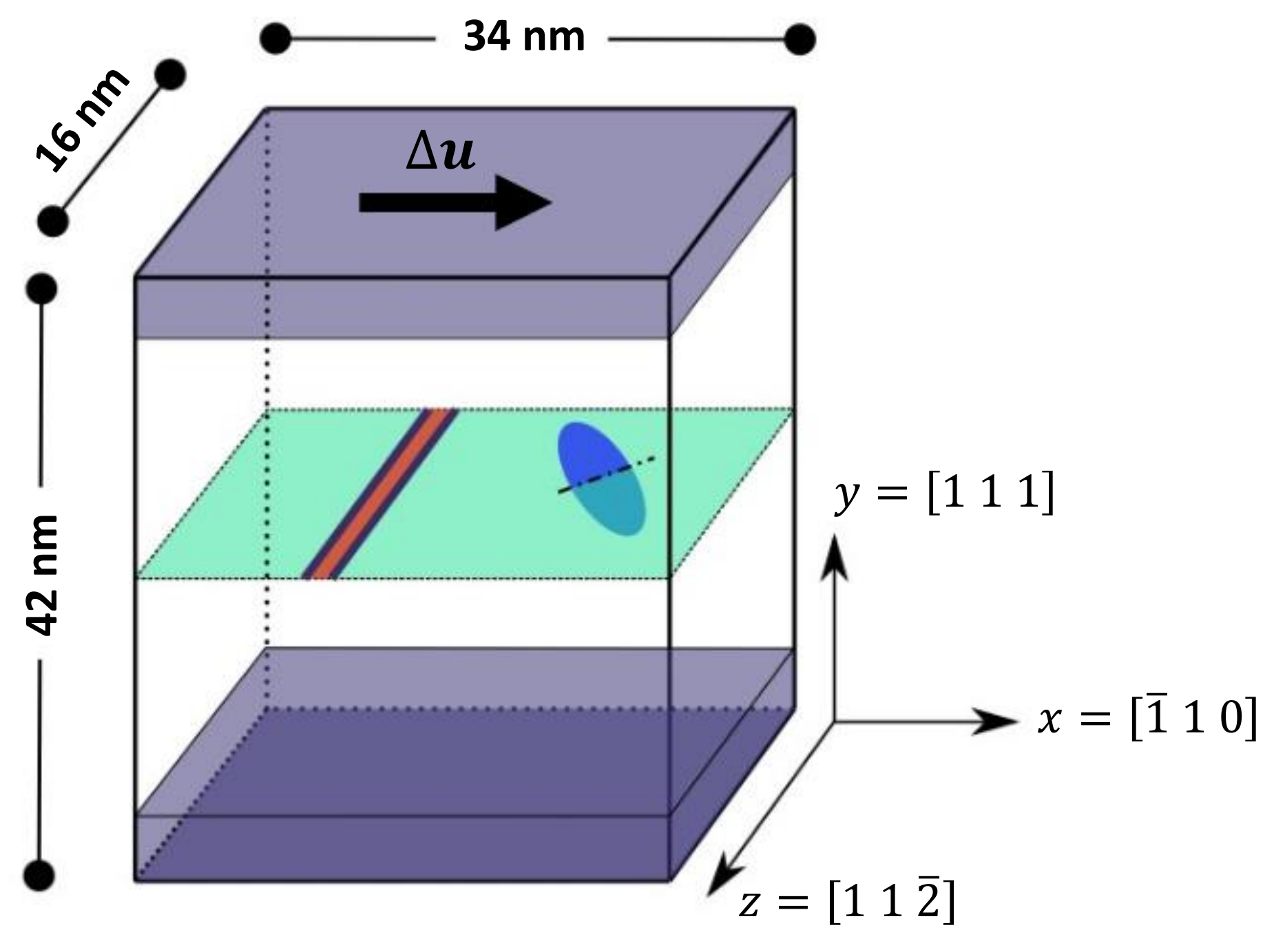} 
\caption{Schematic representation of the parallelepipedic domain used to study the interaction of an edge dislocation with a Guinier-Preston zone in Al-Cu alloy subjected to shear. The domain contains an edge dislocation in equilibrium with a GP zone in a (100) plane. The relative angle between the Burgers vector of the dislocation and the GP zone is 60$^\circ$.}
\label{F1}
\end{figure}

Atomistic simulations of dislocation-precipitate interaction were carried out in a parallelepipedic domain that contains a single dislocation and a GP zone. GP zones are monolayer disks of Cu atoms parallel to the cube faces of the FCC $\alpha$ Al lattice \citep{N14}. FCC crystals are characterized by twelve  $\langle 110\rangle \{111\}$ slip systems, leading to two different geometrical configurations between the dislocation and the GP zone. One of them is shown in Fig. \ref{F1} and will be used for the atomistic simulations. It includes an edge dislocation (dissociated in the leading and trailing Shockley partials) in mechanical equilibrium with a GP zone. The section of the GP along the glide plane forms an angle of 60$^\circ$ with the Burgers vector of a perfect edge dislocation in the FCC lattice. The $X$, $Y$ and $Z$ axes of the domain were parallel to the $[\bar{1}10]$ (slip direction),  $[111]$ (normal to the glide plane) and $[11\bar{2}]$  (dislocation tangent) orientations of the FCC lattice. The dimensions of the domain were 34 $\times$ 42 $\times$ 16 nm$^{3}$. They were chosen following the results of \cite{SC15} to minimize the image stresses in the simulation box.

The edge dislocation was introduced in the domain by applying the corresponding Volterra's displacement field in the case of an isotropic elastic medium \citep{hirel2015atomsk}. After the dislocation was introduced, periodic boundary conditions were applied along the $X$ and $Y$ directions and the system was relaxed using the conjugate gradient algorithm at constant volume, followed by another relaxation at constant pressure to reach the minimum energy configuration and the dislocation was dissociated into two Shockley partials. A GP zone of 4 nm in diameter was  then introduced by substituting the corresponding layer of Al atoms by Cu atoms and the energy of the system was minimized again, leading to an equilibrium configuration between the dislocation and the GP zone. This model stands for a  periodic array of dislocations and precipitates in the slip plane \citep{osetsky2003atomicI}. Shear displacements were applied on the top surface of the domain, parallel to the slip plane. To this end, five layers of atoms at the bottom surface remained fixed during the simulations while five layers of atoms at the top surface were constrained to one-dimensional displacements towards the slip direction (Fig. \ref{F1}).

All atomistic simulations outline in the article have been carried out using LAMMPS \citep{plimpton2007lammps} while OVITO was used to visualize the results \citep{stukowski2009visualization}. All calculations were carried out with the angular dependent potential for the Al-Cu system in \cite{apostol2011interatomic}. This potential follows the embedded atom method scheme  of \cite{daw1984embedded}, with the addition of dipole and quadrupole distortion terms.

Three different types of atomistic simulations were carried out to obtain the parameters needed to determine the rate according to the TST. The first one aimed at exploring the energy landscape of the dislocation-precipitate interaction as a function of the applied strain so the atomic arrangements before and after the dislocation has overcome the GP zone do correspond to minimum energy configurations. This is critical to accurately determine the energy barriers in the following steps but -- due to the presence of many local energy minima in phase space -- it cannot be achieved using standard energy minimization strategies. The second type of simulations use the equilibrium configurations at each strain level to determine the activation enthalpy using a multi-replica optimization procedure \citep{ZLY13}. Finally, standard MD simulations at different strain states and temperatures are carried out to compute the rate at which the dislocation overcomes the GP zone. The details of each methodology are detailed below.

\subsection{Exploration of the energy landscape} \label{ssec3.1:DetermEnergy}

\begin{figure}[h]
\centering
\includegraphics[scale=0.28]{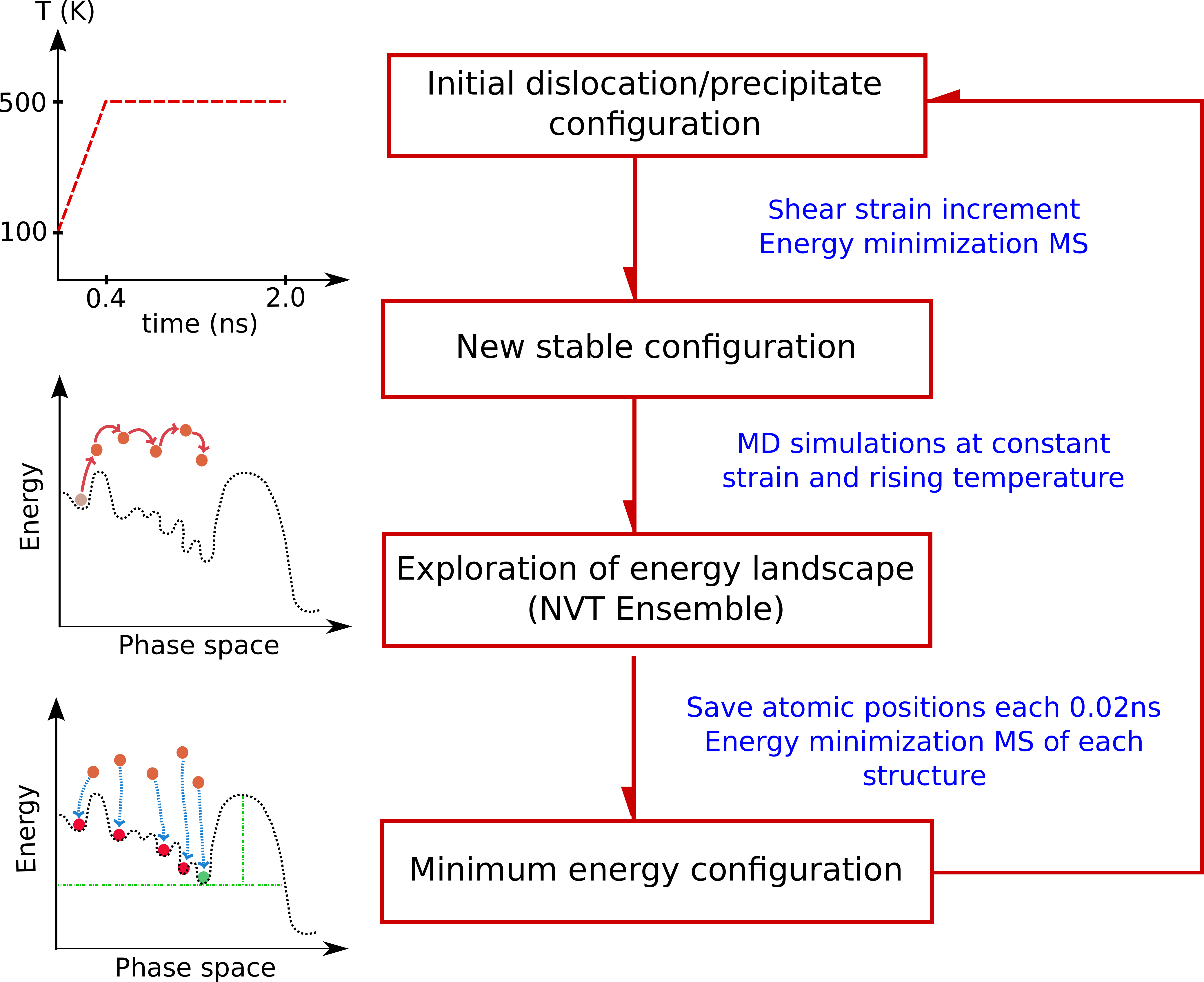} 
\caption{Flowchart of the iterative procedure to explore the energy landscape and obtain the minimum energy configurations in the phase space considered.  Thermal annealing was used to explore different states of the energy landscape, overcoming small transition barriers without changing the energy landscape. This process is illustrated by the energy versus phase space plots. }
\label{F2}
\end{figure}

The energy landscape of the edge dislocation-GP zone interaction presents many local minima. Standard energy minimization procedures based on the application of the conjugate gradient algorithm does not ensure that global minima are attained before and after the dislocation has overcome the precipitate. This limitation was partially mitigated by the application of an iterative strategy that combines MS simulations with thermal annealing using MD. The procedure is detailed in  Fig. \ref{F2}. Starting from the equilibrium configuration detailed above, a shear strain is applied to the simulation domain by applying displacement of 0.25 $\mathrm{\AA}$ to the upper fixed region, followed by an energy minimization using the conjugate gradient algorithm. Then, the potential energy surface at this strain was explored via thermal annealing using MD. To this end, a linear temperature ramp from 100K to 500K was applied in 0.4 ns using a Langevin thermostat within the  canonical ensemble (NVT) and the temperature was held constant at 500 K during 2 ns. Configurations were stored each 0.02 ns and their energy was minimized. A negligible decrease in stress was observed between the directly minimized configuration and after thermal annealing. The configuration with the minimum energy was selected and it was used as the initial structure for the next strain increment.  This thermal annealing procedure to obtain the minimum energy configuration is obviously dependent on the annealing temperature. The energy landscape will not be sufficiently explored if the annealing temperature is too low and the minimum energy configuration will not be reached. On the other hand,  the dislocation will completely overcome the obstacle if the temperature is too high \citep{argon2008strengthening}. Thus, the annealing temperature (and time) was selected by a trial-and-error approach for this particular configuration.  A temperature of 400K was initially chosen but it was not enough to overcome the  local maxima  after an annealing time of 4 ns. So, the annealing temperature was increased up to 500K and 600K and a much more complete exploration of the local minima of the landscape was achieved at these temperatures. So, a temperature of 500K and 4 ns of annealing time was chosen to ensure that high annealing temperatures did not bias the sampling procedure. 
 
\subsection{Determination of the activation energy}

The activation energy $\Delta U(\gamma)$ was computed as a function of the applied shear stress using the NEB method \citep{jonsson1998nudged, henkelman2000climbing}. This multi-replica approach starts from an initial ($R_0$) and final ($R_N$) configuration of the atoms and searches for the minimum energy path between both states using $N$ replicas of the system, which are generated by linear interpolation of the atom coordinates between $R_0$ and $R_N$. These $N$ replicas are connected by harmonic springs to the corresponding atom in the previous ($R_{j-1}$) and the following ($R_{j+1}$) replicas. 

The initial configuration for each value of the applied strain ($R_0 (\gamma)$) was given by the minimum energy configuration at each stress obtained with the thermal annealing procedure detailed above. The final atomic configuration at each strain ($R_N (\gamma)$), was obtained using MD simulations. To this end, the initial atomic configuration $R_0 (\gamma)$ was held at 600 K until the dislocation overcomes the obstacle using the NVT ensemble and the atomic configuration after this event was taken as $R_N (\gamma)$ after minimization. The temperature of 600K was selected so the dislocation bypasses the GP zone in a reasonable time, affordable for MD simulations. It should be noted that the bypass mechanism (shearing of the GP zone by the dislocation) was identical to the one determined in the MS simulations combined with thermal annealing which were detailed above. The fast inertial relaxation engine (FIRE) \citep{bitzek2006structural} with a timestep of 0.01 ps was used along with a spring constant of 10 eV/$\mathrm{\AA}^2$.

Each NEB simulation was carried out with sixteen replicas and the intermediate replicas were obtained by linear interpolation between the initial and the final atomic positions. Nevertheless, it has been shown that this strategy may lead to transition paths that are different from the minimum energy path in complex energetic landscapes because one or more intermediate replicas may get stuck in a local minimum along the elastic band \citep{W11}. Thus, the basin hopping global optimization strategy \citep{SS2014} was also applied within the NEB framework to determine the minimum energy path. In this strategy, a perturbation in the form of small random displacements of the atoms is introduced in each replica along the energetic path, originally obtained through linear interpolation. Each replica is perturbed independently, i.e., atomic displacements were introduced using a different random number seed. The maximum random displacement of the perturbation was 0.08 $\mathrm{\AA}$ in each direction. Then, the minimum energy path is calculated using FIRE, as in traditional NEB, but starting from a perturbed condition at each intermediate replica. Four independent perturbations for each strain were carried out using this BH-NEB methodology and the one showing the minimum energy barrier was accepted as the minimum energy path.

\subsection{Determination of the rate}

The rate at which dislocations overcome the GP zone was determined through MD simulations as a function of the applied strain and temperature. The minimum energy configurations for a given strain obtained as indicated in section \ref{ssec3.1:DetermEnergy} were used as the starting point. MD simulations at constant shear strain were carried out using an NVT ensemble at 400, 450, 500, 550 and 600 K and the time needed for the dislocation to overcome the precipitate, $t^s$, was determined for each case.  Hydrostatic stresses were present in the NVT ensembles at finite temperatures due to thermal expansion but their influence on the rate at which the edge dislocation overcomes the precipitate is negligible because it is the shear stress that controls the driving force. To validate this hypothesis, MD simulations were carried out using a slightly different strategy. The temperature was initially increased from 0 K using an NPT ensemble and the volume was changed to reach the simulation temperature with negligible pressure. Afterwards, simulations were run during 100 ps within the NPT ensemble. Then, an NVT ensemble was used to perform the MD simulation at a given value of the applied strain until the dislocation overcame the GP zone to determine the rate. The rates obtained with this procedure were equivalent to those found using just the NVT ensemble.
 Eight independent simulations were carried out for each strain and temperature to account for the stochastic nature of the process and the average rate, $\overline \Gamma$, was given by $8/\sum t^s_i$.

\section{Results and discussion} \label{sec4:Therm}

\subsection{CRSS and dislocation-GP interaction mechanisms}

\begin{figure}[h]
\centering
\includegraphics[scale=0.8]{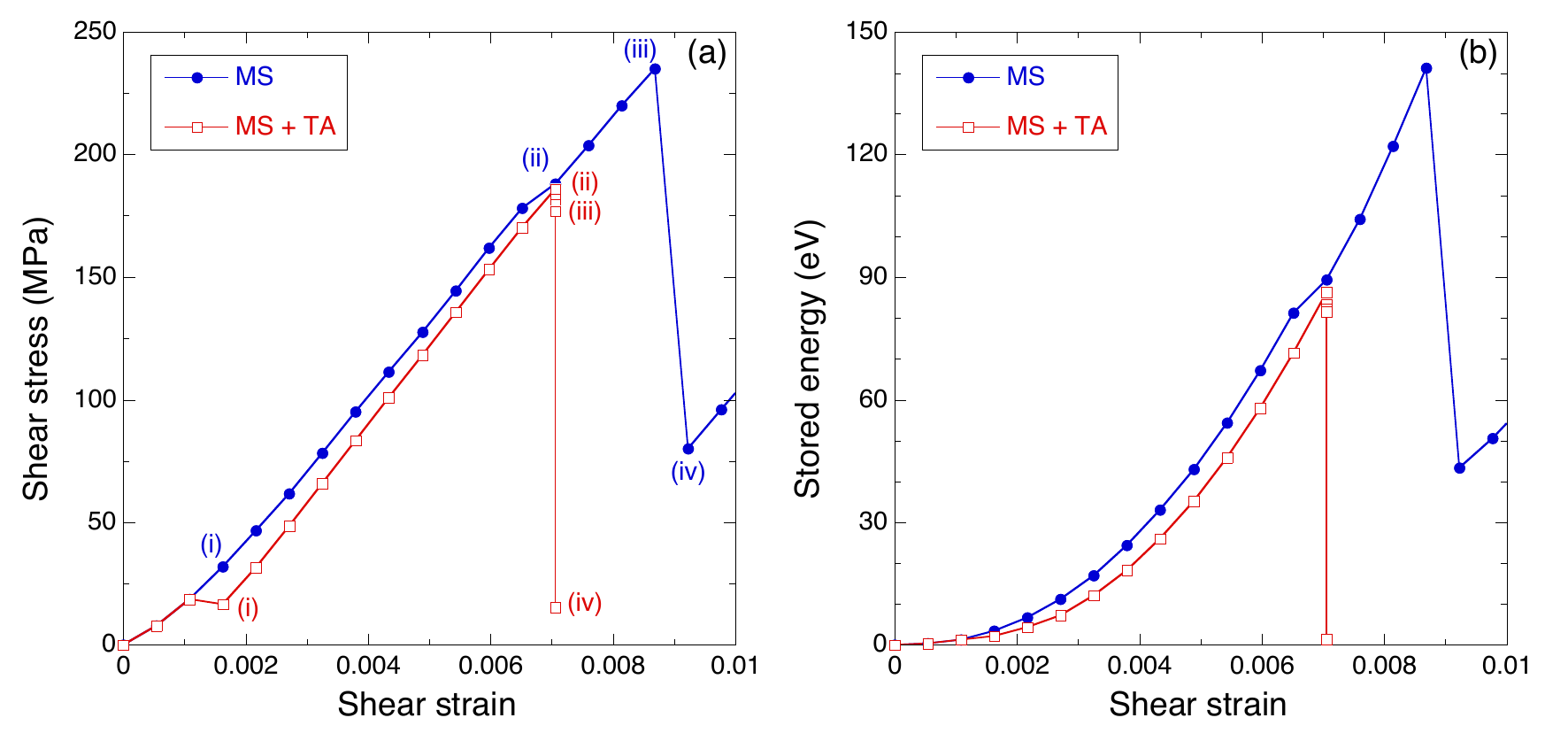} 
\caption{(a) Shear  stress {\it vs.} shear strain curve. (b) Energy stored {\it vs.} shear strain curve. The results were obtained with MS simulations at increasing strain levels (MS) or by MS simulations followed by thermal annealing after each strain increment (MS+TA).}
\label{SE}
\end{figure}

The strategy based on MS simulations at different strains followed by thermal annealing, detailed in section \ref{ssec3.1:DetermEnergy}, provides the shear stress - shear strain behavior of the domain until the dislocation overcomes the GP zone, and it is presented in Fig. \ref{SE}(a). The corresponding curve obtained using only MS (the energy is minimized after each strain increment using the conjugate gradient without any thermal annealing) is also plotted in this figure for comparison. The increment of the energy stored in the domain, $\Delta E$, is plotted in Fig. \ref{SE}(b) as a function of the applied strain in both situations. Although the influence of thermal annealing in the stress-strain and energy stored-strain curves is limited, its effect on the CRSS to overcome the precipitate and on the energy stored at this point was significantly larger. Moreover, this change was associated with a modification of the bypass mechanism.

\begin{figure}[!]
\includegraphics[scale=0.1]{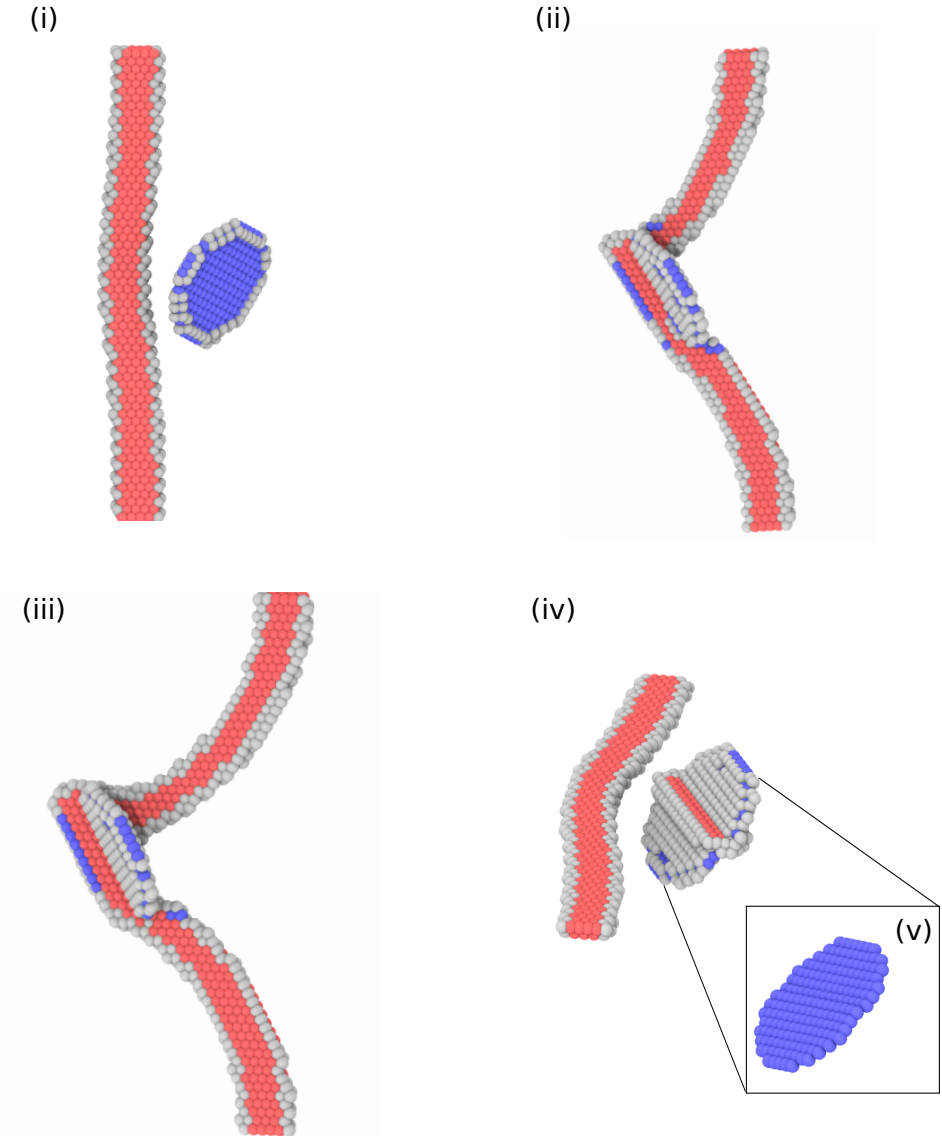} 
\caption{Mechanisms of dislocation-GP zone interaction as a function of the applied shear strain, $\gamma$,  according to MS simulations. Figures (i), (ii), (iii) and (iv) correspond to the points marked with the same letters in blue in Fig. \ref{SE}(a). Figure (v) depicts the state of the GP zone after dislocation bypassing. Cu atoms are shown blue, Al atoms in HCP positions in red and in other positions in white. Al atoms in FCC positions are not shown.}
\label{MS}
\end{figure}

MS simulations without thermal annealing showed that the applied stress had always to increase to move the dislocation closer to the GP zone from the initial equilibrium condition, which is associated with a repulsion between the dislocation and the GP zone. The outward bowing of the dislocation observed in Fig. \ref{MS}(i) is a consequence of such repulsion. When the applied shear stress reached $\approx$ 190 MPa, the leading partial sheared the GP zone, Fig. \ref{MS}(ii), and further deformation led to the formation of an Orowan loop around the GP zone by the trailing partial, as shown in Figs. \ref{MS}(iii) and (iv). The critical shear stress and strain to overcome the GP zone were 235 MPa  and 0.87\% , respectively (Fig. \ref{SE}(a)). The GP zone, also depicted in Figs. \ref{MS}(v), was sheared by the leading partial, creating a step at the center. Both the CRSS and the physical mechanism of dislocation-GP zone interaction are in agreement with the simulations of \cite{SW10} using the same methodology. However, experimental observations find GP zones in Al-Cu to be shearable by dislocations \cite{N14, BFK61, BL18}. Furthermore, GP zone shearing was also reported by MD simulations at 300K in this system \citep{yanilkin2014dynamics}. 

\begin{figure}[!]
\includegraphics[scale=0.1]{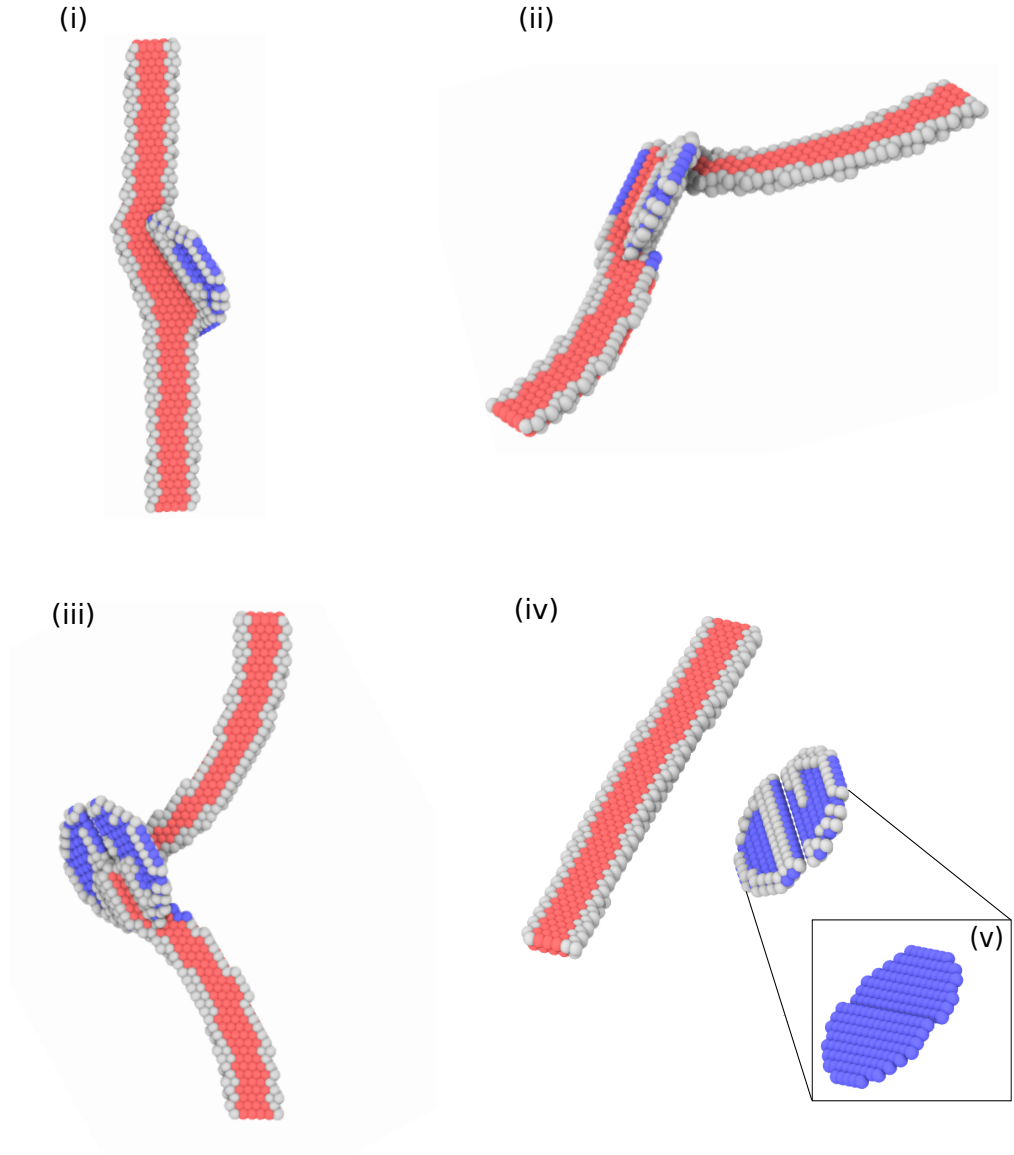} 
\caption{Mechanisms of dislocation-GP zone interaction as a function of the applied shear strain, $\gamma$,  according to MS simulations followed by thermal annealing. Figures (i), (ii), (iii) and (iv) correspond correspond to the points marked with the same letters in red in Fig. \ref{SE}(a). Figure (v) depicts the state of the GP zone after dislocation shearing. Cu atoms are shown blue, Al atoms in HCP positions in red and in other positions in white. Al atoms in FCC positions are not shown.}
\label{MSTA}
\end{figure}

The MS simulations followed by thermal annealing after each strain increment showed subtle differences with the MS results in the initial stages of deformation. The stress initially increased with the applied strain as the dislocation approached the GP zone but then the dislocation was attracted by the GP zone. There was a slight reduction in the shear stress, indicating that the system was slightly overdriven, and the leading partial touched the GP zone, as shown in Fig. \ref{MSTA}(i).  Afterwards, the leading partial progressively sheared the GP zone as the applied strain increased. The critical shear stress and strain to overcome the GP zone were 186 MPa  and 0.71\% , respectively, significantly lower than the ones obtained with MS. The dislocation-GP zone configuration at this point is depicted in Fig. \ref{MSTA}(ii), where the trailing partial is about to start shearing the GP zone. This occurs rapidly afterwards, Fig. \ref{MSTA}(iii), while the shear stress decreases at almost constant strain. Finally, the GP was completely sheared by the dislocation, as shown in Figs. \ref{MSTA}(iv) and (v). 

Similarly, the maximum energy stored in the system just before the dislocation shears the GP zone and after shearing are smaller than that predicted by MS. These results indicate that the energy minima resulting from MS alone probably comes from metastable states due to the rough energy landscape. As a result, the CRSS is overestimated by 20\%  by the MS simulations and more important the initial and final atomic configurations to compute the free energy barrier are different from the ones obtained after a careful search of the energy minima through thermal annealing. This effect leads to different physical mechanisms (Orowan looping {\it vs.} shearing) and to errors  to determine the activation energy through the NEB method \citep{ZLY13}.

It should be noticed that the MS simulations in Fig. \ref{MSTA}(a) depict two different regimes during the dislocation-GP zone interaction. The first one encompasses from the beginning of the stress-strain (or energy-strain) curve until point (i) in Fig. \ref{MSTA}(a). Within this range, the dislocation is approaching the precipitate and there are elastic and plastic contributions to the deformation. The second regime covers from point (i) to point (ii) in Fig. \ref{MSTA}(a). In this range,  the dislocation is in contact with the GP zone and there is only elastic contribution to the deformation. Thus, the slope of the stress-strain curve equals the shear modulus of the Al alloy in the local coordinate system, $\mu'$, which is equal to  31.6 GPa. This linear relationship between the shear stress and strain allows to relate the activation Helmhotz free energy at constant strain, $\Delta F(\gamma)$,  with the activation Gibbs free energy at constant stress, $\Delta G(\tau)$, as explained in section \ref{sec2:TST}.

\subsection{Activation Helmholtz free energy and activation energy}

\begin{figure}[!]
\centering
\includegraphics[scale=0.75]{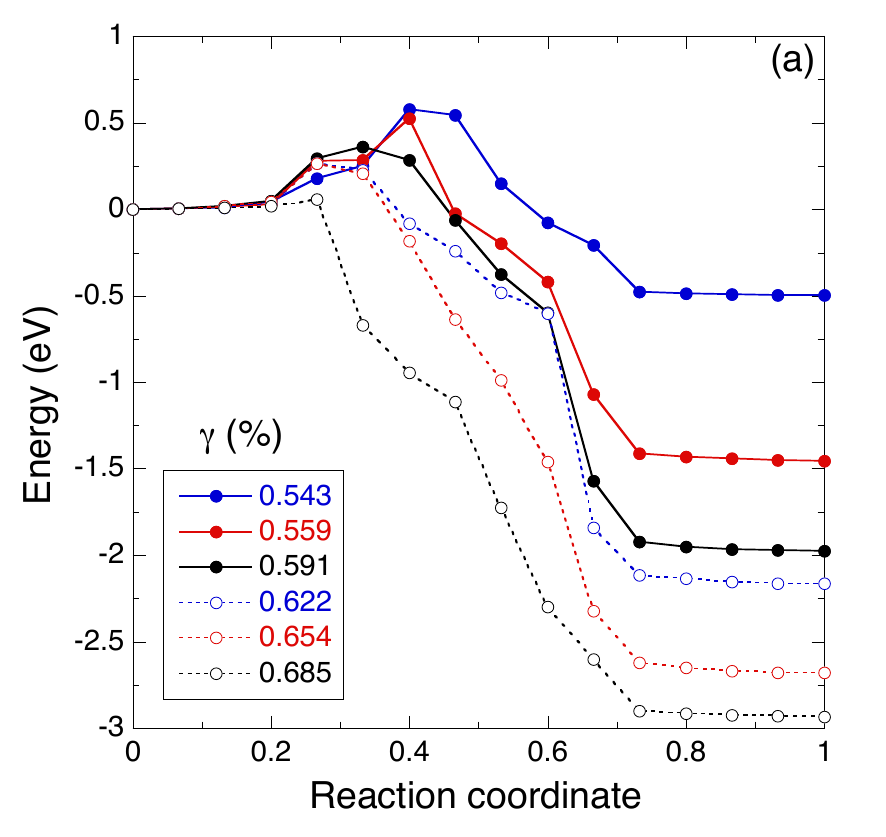} 
\includegraphics[scale=0.75]{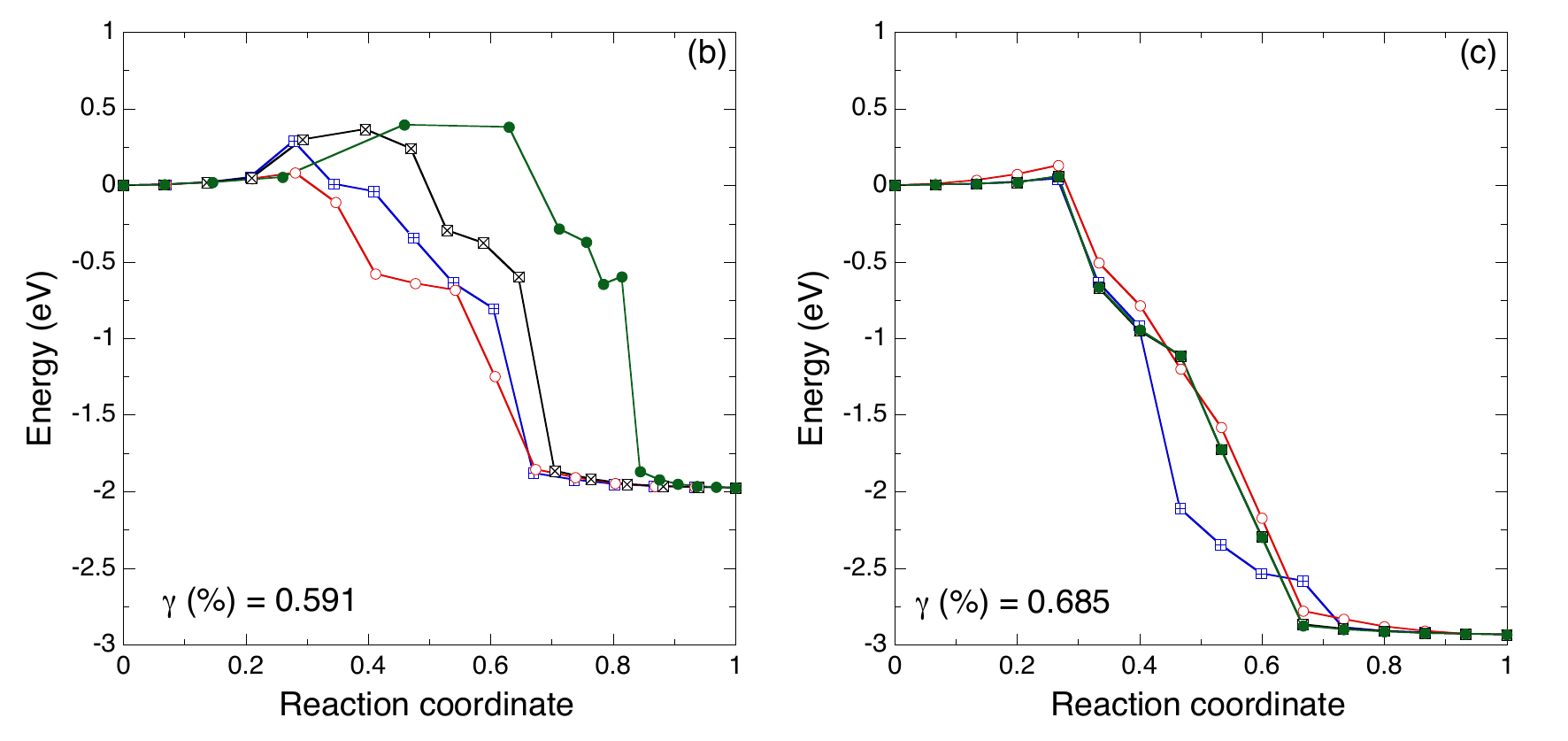} 
\includegraphics[scale=0.75]{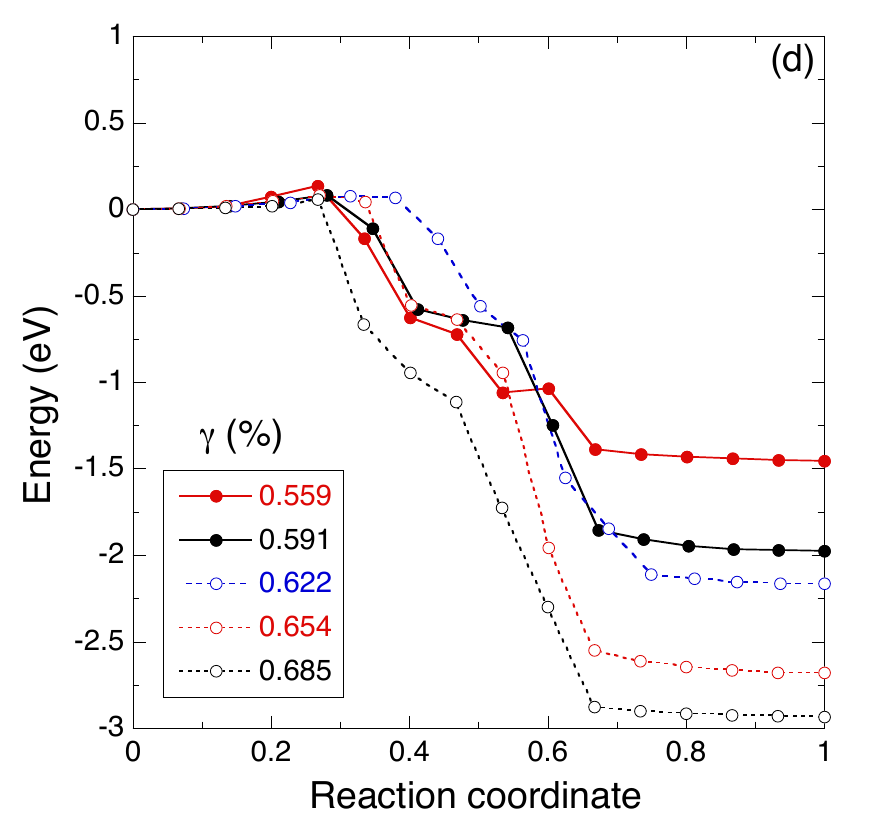} 
\caption{Evolution of the energy of the system along the reaction path as a function of the applied shear strain. (a) Results of standard NEB calculations with linear interpolation between basins. (b) and (c) Results of the basin hopping global optimization NEB calculations for applied strains of 0.591\% and 0.685\%, respectively. Four different transition paths were obtained in (b) and (c) for each strain as a function of the initial random perturbation. (d) Minimum energy paths according to the basin hopping NEB calculations.}
\label{NEB}
\end{figure}

The activation energy $\Delta U(\gamma)$ was evaluated using the standard NEB and the basin hopping global optimization methodology (NEB-BH). The results of the calculations obtained with the standard NEB method are shown in Fig. \ref{NEB}(a) for different applied strains. The saddle points of the NEB calculations depicted in this figure provide the evolution of the energy barrier, $\Delta U$, with strain. It is worth mentioning that the final configuration is indeed an intermediate minimum but the barrier to reach the final minimum is one to two orders of magnitude smaller than the barrier shown, which becomes the rate limiting step.

In the case of the basin hopping global optimization, the four random perturbation applied led to different transition paths, which are plotted in Figs. \ref{NEB}(b) and (c) for applied shear strains of 0.591\% and 0.685\%. Similar results were obtained for other values of the applied strain and are not included for the sake of brevity. Large differences in the energy barrier were found  for the different transition paths and the energy barriers corresponding to the minimum energy paths shown in Figs. \ref{NEB}(b) and (c) were much lower than those found following the standard NEB method (Fig. \ref{NEB}(a)). The transition paths that led to the minimum energy barrier for different values of the applied strain $\gamma$ according to the basin hopping  global optimization are plotted in Fig. \ref{NEB}(d). The differences with those obtained with the standard NEB minimization in Fig. \ref{NEB}(a) are very large and these results indicate -- in agreement with previous investigations \citep{W11} -- that the standard NEB method with linear interpolation is not able to provide an accurate estimation of the minimum energy path due to the roughness of the energy landscape.

\begin{figure}[!]
\centering
\includegraphics[scale=0.8]{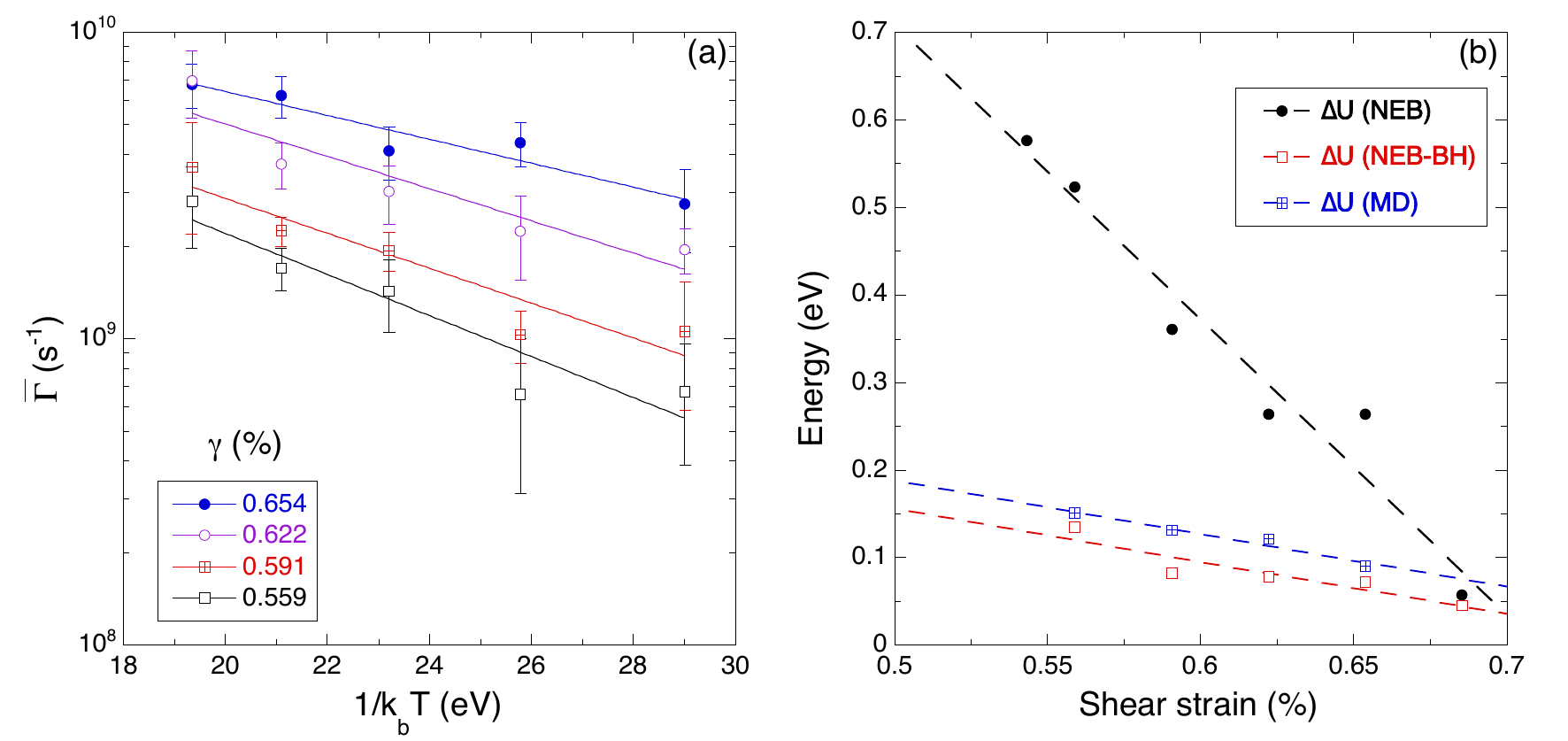}
\caption{(a) Average rate, $\overline \Gamma$, for the dislocation to overcome the GP zones as function of $1/k_bT$ for different applied strains. The slope of each curve represents $\Delta U(\gamma)$. (b) Variation of the activation energy, $\Delta U$, as a function of the applied shear strain $\gamma$ obtained by different atomistic techniques. See text for details.}
\label{dU}
\end{figure}

The activation Helmholtz free energy $\Delta F(\gamma)$ was obtained using direct MD simulations. These calculations, as explained in section \ref{sec3:Proc}, provide the average rate, $\overline \Gamma$, for the dislocation-GP interaction  as a function of the temperature and the applied strain  and the results are plotted in Fig. \ref{dU}(a). It can be noted that the rate at a given strain can be fit with an Arrhenius equation (Eq. \ref{E3}). Thus, it is confirmed that $\Delta F$ is a function of the applied strain.  In the case that the HTST is applicable, the entropy of the system does not depend on temperature, and the slopes of the $\overline \Gamma$ {\it vs.}  1/$k_bT$ lines provide the activation energies $\Delta U(\gamma)$, while the effective attempt frequency ($\nu_{eff}$) in eq. (\ref{E7}) (which contains the activation entropy contribution, $\Delta S_v$) is given at the limit when 1/$k_bT$ $\rightarrow 0$.

The activation energies obtained with the different atomistic techniques (molecular dynamics, standard NEB  and basin hopping NEB) are plotted in Fig. \ref{dU}(b) as a function of applied the shear strain. The activation energies decrease linearly with the strain in all cases and the values of the activation energy obtained through the MD simulations and the basin hopping NEB are in very good agreement, taking into account the scatter associated with both types of simulations. This result indicates that entropic contribution to the rate is independent of the temperature and strain and HTST can be applied to study the interaction of edge dislocations with GP zones. It should also be noticed that the standard NEB method leads to transition paths which are far away from the minimum energy path and that the differences increased rapidly as the strain decreased. Thus, this methodology is not appropriate to determine the activation energy due to the roughness of the energy landscape. 

Kocks \cite{kocks1975thermodynamics} proposed a phenomenological model for the activation free energy  $\Delta G(\tau)$ to overcome an array of obstacles in the dislocation glide plane  
\begin{equation}
\Delta G(\tau)= \Delta F_0\left[1-\left(\frac{\tau}{\tau_{0}}\right)^{p}\right]^q
\label{E8}
\end{equation} 

\noindent where $\Delta F_0$ denotes the obstacle energy barrier at zero stress (or strain), $\tau_0$ is the CRSS to overcome the obstacle at 0K, and $p$ (0 $< p < 1$) and $q$ (1 $< q < 2$) are two parameters which depend on the randomness of the obstacle distribution in the glide plane and on the potential energy landscape of the interaction, respectively. For a perfect array of aligned precipitates the force exerted by the obstacle against the dislocation is constant and equal to $\tau b L$, where $L$ is the distance between obstacles along the dislocation line. Under these conditions, and taking into account that the precipitates form a regular array in our atomistic simulations, it is easy to show that $p$ = 1 \citep{Hull2001,caillard2003thermally}. Therefore,

\begin{equation}
\Delta G (\tau)= \Delta F_0\left(1- \frac{\tau}{\tau_{0}}\right)^q
\label{E9}
\end{equation}

Similarly, the dependence of the activation energy, $\Delta U$, with the applied shear strain $\gamma$ obtained from MD simulations could be approximated by a linear function of $\gamma$ in the same form as  eq. \eqref{E9}

\begin{equation}
\Delta U (\gamma) =  \Delta U_0 \left(1-\frac{\gamma}{\gamma_0}\right)
\label{E10}
\end{equation}

\noindent where $\Delta U_0$ and $\gamma_0$ stand for the activation energy at zero strain and the critical strain to overcome the obstacle at 0K, which are depicted in Table \ref{FitParam}.  
\begin{table}[h]
\begin{center}
\caption{Parameters to relate the activation energy  barrier with the applied strain, eq. (\ref{E10}).} 
\begin{tabular}{c c c}
 \hline
 $\Delta U_0$ (eV) & $\gamma_0$ ($\%$) \\
  \hline
  0.45 & 0.80  \\
  \hline
\end{tabular}
\label{FitParam}
\end{center}
\end{table}

\subsection{Activation entropy}

The activation entropy $\Delta S_\gamma$ was obtained  for the different values to the applied strain from a effective attempt frequency $\nu_{eff}$, which was determined from  the MD simulations (eq. \eqref{E7}), assuming a fundamental attempt frequency $\nu \approx 10^{-10}$ Hz. \cite{MR04, sobie2017modal}. $\Delta S_\gamma$ was in the range 1.3-1.8$k_b$ and did not seem to depend on the strain, taking into account the scatter in the rates obtained by MD. The values of $\Delta S_\gamma$ are in agreement with the entropic contribution associated with the typical vibrational entropy of solids and are similar to those reported for the dislocation nucleation \cite{HL10, SR11}. Thus, this result supports the applicability of the HTST to analyze the dislocation/GP zone interaction in Al-Cu alloys. It should be noted, however, that Saroukhani {\it et al.} \cite{saroukhani2016harnessing} estimated an activation entropy of  13.2$k_b$ for the same process. This large value of the entropy was associated to anharmonic effects induced by thermal softening, which were accounted for by means of the Meyer-Neldel rule \citep{K73}. One important difference between the MD simulations of Saroukhani {\it et al.} \cite{saroukhani2016harnessing} and our simulations is the size of the simulation domain, which only contained 13000 atoms in the former. The results obtained with such small domain may be strongly influenced by image forces \cite{SC15}, leading to important differences in the rate at which the dislocation overcomes the GP zone. In addition, the MD simulations in \cite{saroukhani2016harnessing}
were carried out under constant strain and, as indicated in eq. \eqref{E5}, the entropic contributions under applied stress and strain can be very different if the volume of the crystal is small.

\subsection{Activation volume}

The activation volume is defined as

\begin{equation}
V_0=\left. - \frac{\partial \Delta G (\tau)}{\partial \tau} \right|_T
\label{E11}
\end{equation}

\noindent but the MD simulations presented above were carried out at constant strain $\gamma$ and provided the value of $\Delta F(\gamma)$ instead of $\Delta G(\tau)$. Nevertheless, the relationship between both magnitudes can be obtained following Ryu {\it et al.} \cite{SR11}, as detailed in the Appendix, and it is given by

\begin{equation}
\Delta F (\gamma) = \Delta G(\tau_m) - \frac{1}{2} V \frac{(\tau_s-\tau_m)^2}{\mu'}
\label{E12}
\end{equation}

\noindent where $\tau_m$ is the applied stress at the local minimum, which corresponds to  point (ii) in Fig. \ref{MSTA}(a), $\tau_s$ is the stress in the saddle point, which corresponds to point (iii) in Fig. \ref{MSTA}(a), and $V$ is the volume of the crystal. Finally, $\mu'$ is the local shear modulus of the crystal that defines the slope of the $\tau$-$\gamma$ linear relationship between points (i) and (ii) in Fig. \ref{MSTA}(a). In the case of the interaction between the dislocation and the GP zone, the difference $\tau_s-\tau_m \approx$  1 MPa and the second term in eq. \eqref{E15} is negligible. Thus, $\Delta F(\gamma)$ $\approx$ $\Delta G(\tau)$ and the activation volume can be estimated in this case, following eq. \eqref{E12}, as  

\begin{equation}
V_0 \approx \left. -\frac{\partial (\Delta F (\gamma))}{\partial \tau} \right|_T
= \left. - \frac{1}{\mu'} \frac{\partial (\Delta F(\gamma))}{\partial \gamma} \right|_T
\label{E13}
\end{equation}

\noindent  leading to $V_0$  $\sim 12b^3$. This magnitude is in good agreement with previously reported values \citep{Asaro2005, Wang2005, Zhu2008} and supports the hypothesis that $\Delta F(\gamma)$ $\approx$ $\Delta G(\tau)$ in this case as $V_0 << V$.

\section{Predictions of the flow stress} \label{sec5:FlowStress}

An estimation of the flow stress of Al alloys containing GP zones as a function of temperature can be obtained relying on the TST. The HTST rate can be related to the plastic strain rate, $\dot\gamma_p$, through the well-stablished model proposed by Kocks \cite{kocks1975thermodynamics}

\begin{equation}
\dot\gamma_p = \dot\gamma_{p0} \exp{\bigg(\frac{-\Delta G(\tau)}{k_bT}\bigg)} = \dot\gamma_{p,eff} \exp{\bigg(\frac{-\Delta H(\tau)}{k_bT}\bigg)}
\label{E14}
\end{equation}

\noindent where $\dot\gamma_{p, eff}$ stands for the reference plastic strain rate, that can be expressed as \cite{O34},

\begin{equation}
\dot\gamma_{p, eff} = \rho b L \nu_{eff}
\label{E15}
\end{equation}

\noindent where $L$ is the mean free path of the dislocation (the distance between GP zones in the slip direction, $L_x$ in our simulations), $\nu_{eff} =\nu \exp{\left(\Delta S_\gamma/k_b\right)}$  the fundamental attempt frequency (= 0.4-0.5 $\times$ 10$^{11}$ s$^{-1}$ according to the direct MD simulations above) and $\rho$ stands for the dislocation density. The actual dislocation density in the MD simulations is very high and the predictions of the flow stress of should be carried out using dislocation densities typical of a well-annealed crystal, $2 \times 10^{11}$ m$^{-2}$ \citep{HSL18}.

When HTST holds, Kocks' model can be expressed according to 

\begin{equation}
\Delta H (\tau) = \Delta H_0 \left(1-\frac{\tau}{\tau_0}\right)
\label{E16}
\end{equation}

\noindent where $\Delta H_0$ is the total barrier to overcome in absence of stress and temperature \citep{Hara2010}. Thus, the yield stress $\tau_y$ can be expressed, according to eqs. \eqref{E14} and \eqref{E16}, as

\begin{equation}
\tau_y =  \tau_0 \bigg[ 1- \frac{k_b T}{\Delta H_0}   \ln \left(\frac{\dot\gamma_{eff}}{\dot\gamma}\right)\bigg].
\label{E17}
\end{equation}
\noindent where $\tau_0$ is the stress to overcome the barrier without any thermal contribution.   

As explained above, in the limit $V > > V_0$ and $\tau=f(\gamma)$, it can be assumed that $\Delta G(\tau) \approx \Delta F(\gamma)$, which also implies $\Delta H_0  \approx  \Delta U_0$. Thus, eq. \eqref{E17} can be re-written as

\begin{equation}
\tau_y =  \tau_0 \bigg[ 1- \frac{k_b T}{\Delta U_0}   \ln \left(\frac{\dot\gamma_{eff}}{\dot\gamma}\right)\bigg].
\label{E18}
\end{equation}

\noindent where $\tau_0$ = 186 MPa is the CRSS to overcome the GP at 0K, according to Fig. \ref{MS}(a).

\begin{figure}[h]
\centering
\includegraphics[scale=1]{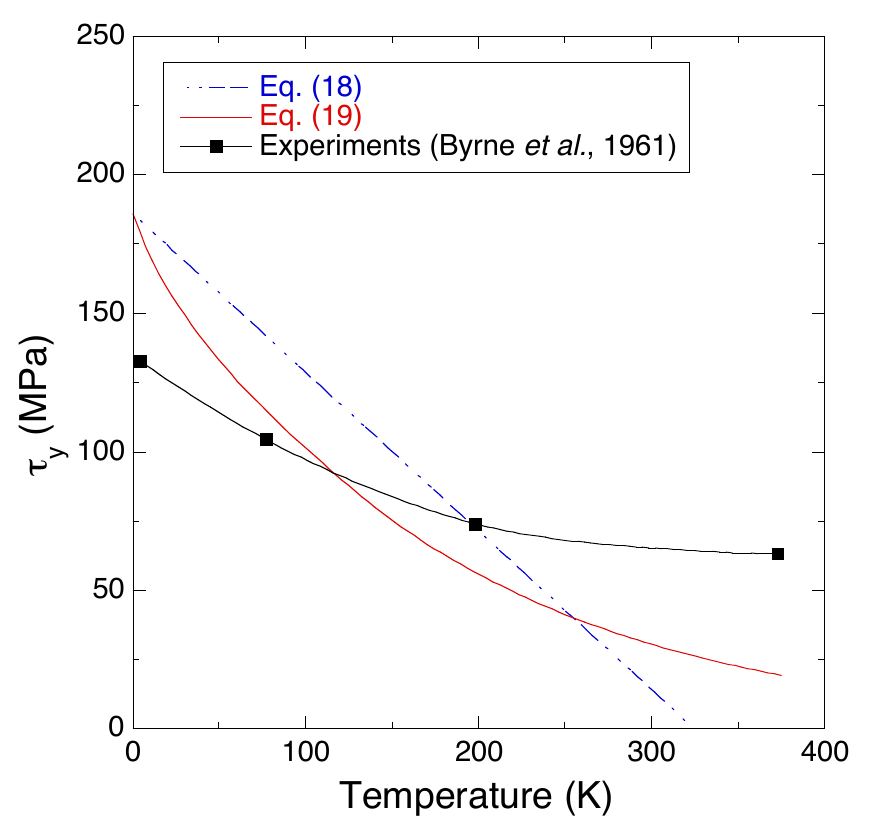}
\caption{Experimental results of the shear flow stress, $\tau_y$ of an Al alloy containing GP zones as a function of temperature \cite{BFK61} and predictions based on TST according to eqs. (\ref{E18}) and (\ref{E19}).}
\label{exp}
\end{figure}

The influence of the temperature on the shear flow stress, $\tau_y$ of an Al - 4 wt. \% Cu alloy containing GP zones was measured by Byrne {\it et al.} \cite{BFK61} between 4 K and 200K in single crystals oriented for single slip. No information was provided in this investigation about the spacing between GP zones and the initial dislocation density but the values used for these parameters in eq. \eqref{E15} were comparable with those measured in an Al - 4 wt. \% Cu alloy aged at ambient temperature \cite{RBP18}. GP zones were sheared by dislocations in the whole temperature range and the experimental values of $\tau_y$ at the onset of yielding are plotted in Fig. \ref{exp} together with the predictions of eq. (\ref{E18}) for a strain rate of 10$^{-2}$ s$^{-1}$. The results obtained from the TST and eq. (\ref{E18}) tend to overestimate the flow stress at 0K and, more importantly, predict a reduction in the flow stress with temperature which is much higher than the experimental one and which leads to a negligible flow stress at 320K.
This latter difference appears because eq. \eqref{E17} only includes the thermal contribution to the flow stress due to the presence of the GP zones and neglects the athermal contribution due to dislocation/dislocation interactions. The athermal effect is particularly important at ambient and elevated temperature and is responsible for the finite value of the experimental flow stress at these temperatures.

The differences between eq \eqref{E18} and the experimental results in the low temperature regime can be attributed to different factors. Firstly,  the precipitates in the experimental samples are randomly distributed  and  recent findings suggest that the Helmholtz activation free energy varies with the distribution of the precipitates in the material  and it is smaller in the case of a random distribution as compared with a  regular distribution \citep{sobie2017thermal}. Thus, different distribution of precipitates could lead to a variation of $\Delta F$, providing different predictions of $\tau_y$. In addition,  the GP zones (which grow parallel to the (001) planes of the FCC lattice) have two different orientations with respect to the glide plane. The angle between the Burgers vector and the section of the precipitate along the glide plane is 0$^\circ$ in one case and 60$^\circ$ in the other case. Our atomistic simulations were carried out in GP zones oriented at 60$^\circ$ (Fig. 1) but GP zones oriented at 0$^\circ$ are softer than those at 60$^\circ$ \cite{SW10}. This may explain the overestimation of the shear flow stress at low temperatures. Finally, plastic deformation is caused by  edge, screw and mixed dislocations but our atomistic simulations only consider an edge dislocation interacting with a periodic array of precipitates oriented at 60$^\circ$.

The effect of the random distribution of the precipitates  can be partially overcome based on the results of Leyson and Curtin \cite{leyson2012B, LC16}, who  proposed  a  model based on the TST for the thermally-activated flow stress of Al  in the presence of a random distribution of obstacles. The shear flow stress, $\tau_y$, can be expressed as \cite{leyson2012B, LC16}

\begin{equation}
\tau_y =
\begin{cases}  
\tau_0 \bigg[ 1- \left[ \frac{k_b T}{\Delta U_0} \ln \left(\frac{\dot\gamma_{eff}}{\dot\gamma} \right) \right]^{\frac{2}{3}} \bigg] & \mbox{if } \:\:\frac{\tau_y(T)}{\tau_0} \ge 0.55 \\
\\
\tau_0 \exp{\bigg[-\frac{1}{0.51}\frac{k_b T}{\Delta U_0} \ln \left(\frac{\dot\gamma_{eff}}{\dot\gamma}\right)\bigg]}, & \mbox{if } \:\:\frac{\tau_y(T)}{\tau_0} < 0.55
\end{cases}
\label{E19}
\end{equation}

The predictions obtained with eq.(\ref{E19}) are also plotted in Fig.\ref{exp} and show better agreement with the experimental data although some differences between predictions and experiments are still present. This implies that that the factors discussed above need be taken into account to make accurate predictions of the strengthening provided by GP zones based on atomistic simulations and TST.

\section{Conclusions} \label{sec6:Conclusion}

The interaction of  edge dislocations with Gunier-Preston zones in Al-Cu alloys has been studied by means of atomistic simulations within the framework of the transition state theory.  The different thermodynamic functions that determine the features of the Guinier-Preston barrier were computed. A strategy based on molecular statics in combination with thermal annealing was used to simulate the mechanism of shearing of the Guinier-Preston zones as well as the initial and final configurations to determine the activation enthalpy through the nudged elastic band method. The activation Helmholtz free energy was determined by means of direct molecular dynamics simulations.

It was found that the rate at which dislocations overcome the precipitate is controlled by the activation energy, $\Delta U$, in agreement with the postulates of harmonic transition state theory. The entropic contribution to the free energy was in the range $\Delta S_\gamma \approx 1.3-1.8k_b$ and did not depend on temperature. Moreover, these values of $\Delta S_\gamma$ are in agreement with the entropic contribution associated with the typical vibrational entropy of solids. The activation volume, $V_0$ = 12$b^3$ was determined from the activation Helmholtz free energy and was in good agreement with previously reported values \citep{Asaro2005, Wang2005, Zhu2008}.

Finally, an estimation of the initial shear flow stress as a function of temperature was carried out based on the Kocks'  model \cite{kocks1975thermodynamics} for thermally-activated plastic flow and the activation Gibbs energy  computed by means of molecular dynamics simulations. Some discrepancies between the experimental results and our predictions were found. 
At ambient and elevated temperature, they were expected because the model did not include the athermal contribution associated with the dislocation/dislocation interactions.  At low temperature,  the model overestimates the strength and this result  could be traced to the effect of the random precipitate distribution in the actual samples (which could affect the Helmholtz activation energy \cite{sobie2017thermal}), to the different orientations between the GP and the dislocations and to the different dislocation characters. The influence of the random precipitate distribution was explored based on the results of another model which takes this effect into account \citep{leyson2012A,leyson2016}, leading to a better agreement with the experimental results.

\section*{Acknowledgements}
This investigation has been supported by the European Research Council under the European
Union's Horizon 2020 research and innovation programme (Advanced Grant VIRMETAL, grant
agreement No. 669141). The authors acknowledge the Supercomputing and Visualization Center of Madrid (CeSViMa) for the computer resources, technical expertise and assistance provided. Additionally, the authors thankfully acknowledge the computer resources at Finisterrae and the technical support provided by CeSGa and Barcelona Supercomputing Center (project QCM-2017-3-0007). Finally, use of the computational resources of the Center for Nanoscale Materials, an Office of Science user facility, supported by the U.S. Department of Energy, Office of Science, Office of Basic Energy Sciences, under Contract No. DE-AC02-06CH11357, is gratefully acknowledged.


\appendix

\section{Relationship between $\Delta G(\tau)$ and $\Delta F(\gamma)$} \label{ApxA}

The relationship between the Gibbs activation free energy at constant stress, $\Delta G(\tau)$ and the Helmholtz activation free energy at constant strain, $\Delta F(\gamma)$ can be derived  following \cite{SR11}. Let us consider our system with a prescribed strain $\gamma$. The Helmholz activation free energy $\Delta F(\gamma)$ is the difference between the Helmholtz free energy between the local minima $F_m(\gamma)$ and the saddle point $F_s(\gamma)$. The Gibbs free energy that corresponds to these two states can be written as

\begin{equation}\label{EA1}
F_m(\gamma)=G_m(\tau_m)+\tau_m \gamma V \qquad {\rm and} \qquad
F_s(\gamma)=G_s(\tau_s)+\tau_s \gamma V
\end{equation}

\noindent where the subindexes $m$ and $s$ denote the local minimum and the saddle point, respectively. It is worth noting that $G$ in the local minimum and the saddle point do not correspond to the same shear stress, which is $\tau_m$ in the minimum and $\tau_s$ in the saddle point. 

The activation Helmholtz free energy can be expressed as a function of the Gibbs free energy as

\begin{equation}
\Delta F (\gamma)= F_s(\gamma)- F_m(\gamma)= G_s(\tau_s)-G_m(\tau_m)+\gamma V(\tau_s-\tau_m)
\label{EA3}
\end{equation}

\noindent and it should be noticed that that $G_s(\tau_s)-G_m(\tau_m)$ is not the activation Gibbs free energy since  $\tau_m \ne \tau_s$. If $G_s(\tau_m)$ is added and subtracted to eq. \eqref{EA3},

 \begin{equation}
\Delta F (\gamma) =  \Delta G(\tau_m) + G_s(\tau_s)-G_s(\tau_m)+\gamma V(\tau_s-\tau_m)
\label{EA4}
\end{equation}

\noindent where $\Delta G(\tau_m) = G_s(\tau_m)-G_m(\tau_m)$.  

Expanding $G(\tau)$ around $(\tau_s-\tau_m)$,  asumming $\tau_m$ constant and neglecting the third order terms, eq. \eqref{EA4} can be rearranged as 

 \begin{equation}
\Delta F (\gamma) \approx  \Delta G(\tau_m)+ \gamma V(\tau_s-\tau_m) + \left. \frac{\partial G}{\partial \tau}\right|_V(\tau_s-\tau_m) + \left. \frac{1}{2}\frac{\partial^2 G}{\partial \tau^2}\right|_V (\tau_s-\tau_m)^2.
\label{EA5}
\end{equation}

$\tau$ and $\gamma$  are related to each other according to 

\begin{equation}\label{EAeps}
\left. \gamma = - \frac{1}{V} \frac{\partial G}{\partial \tau}\right|_T 
\end{equation}

\noindent and introducing eq. \eqref{EAeps} in eq. \eqref{EA5}, leads to  

 \begin{equation}\label{EA6}
\left. \Delta F (\gamma) = \Delta G(\tau_m)+ \frac{1}{2}\frac{\partial^2 G}{\partial \tau^2}\right|_V(\tau_s-\tau_m)^2.
\end{equation}

Moreover, differentiating eq. \eqref{EAeps} leads to

\begin{equation}\label{EA7}
\left. \frac{\partial^2 G}{\partial \tau^2}\right|_T = \left. - V \frac{\partial \gamma}{\partial \tau}\right|_T = \frac{-V}{\mu'}
\end{equation}
\noindent where $\mu'$ is the slope of the $\tau$-$\gamma$ curve that describes the elastic interaction of the dislocation with the GP zone before the obstacle is overcome. If eq. \eqref{EA7} is introduced into eq. \eqref{EA6}

 \begin{equation}\label{EA8}
\Delta F (\gamma) = \Delta G(\tau_m)-\frac{1}{2} V \frac{(\tau_s-\tau_m)^2}{\mu'}
\end{equation}

\end{document}